\documentclass[preprint2]{aastex}

\usepackage{amsmath}

\shorttitle{HD 148293 Diameter and Temperature}
\shortauthors{Baines et al.}

\begin{document}

\title{The Angular Diameter and Effective Temperature of the Lithium-Rich K Giant HD 148293 from the CHARA Array}

\author{Ellyn K. Baines}
\affil{Remote Sensing Division, Naval Research Laboratory, 4555 Overlook Avenue SW, \\ Washington, DC 20375}
\email{ellyn.baines@nrl.navy.mil}

\author{Harold A. McAlister, Theo A. ten Brummelaar, Nils~H.~Turner, Judit Sturmann, \\ Laszlo Sturmann, P. J. Goldfinger, Christopher D. Farrington}
\affil{Center for High Angular Resolution Astronomy, Georgia State University, P.O. Box 3969, \\ Atlanta, GA 30302-3969}

\author{Stephen T. Ridgway}
\affil{National Optical Astronomy Observatory, P.O. Box 26732, Tucson, AZ 85726-6732} 

\begin{abstract}
We measured the angular diameter of the lithium-rich K giant star HD 148293 using Georgia State University's Center for High Angular Resolution Astronomy (CHARA) Array interferometer. We used our measurement to calculate the star's effective temperature, which allowed us to place it on an H-R diagram to compare it with other Li-rich giants. Its placement supports the evidence presented by \citet{2000AandA...359..563C} that it is undergoing a brief stage in its evolution where Li is being created.
\end{abstract}


%%%%%%%%%%%%%%%%%%%%%%%%%%%%% Introduction %%%%%%%%%%%%%%%%%%%%%%%%%%%%%
\section{Introduction}
G and K giant stars are expected to have low lithium abundances because Li is destroyed in main-sequence stars except in the outermost layers (1$\%$ - 2$\%$ by mass). As the star ascends the red giant branch of the H-R diagram, the convective envelope deepens, diluting the existing Li and further reducing the observable Li \citep{1989ApJS...71..293B}. This effect is seen in most G and K giant stars, though a small minority ($\sim 1 \%$) exist that are unexpectedly rich in Li \citep[e.g.,][]{1982ApJ...255..577W,1990AJ.....99.1225P,1993ApJ...403..708F}.

Lithium abundance calculations are sensitive to temperature variations, so knowing the effective temperature ($T_{\rm eff}$) of a star is vital in estimating its Li content. HD~148293 was discovered to be Li rich by \citet{1989ApJS...71..293B}, who estimated its temperature using published photometry and color-temperature relations. Their value was also used by \citet[][hereafter CB00]{2000AandA...359..563C}, who placed HD 148293 on an H-R diagram and found it was at a phase known as the ``bump in the luminosity function''. This phase is characterized by an outwardly-moving hydrogen shell, leading to a short-lived phase of Li production before it is rapidly destroyed as evolution continues. Only low-mass stars that contain a highly degenerate helium core and later experience the helium flash pass through this stage and spend a mere 3$\%$ of their ascent on the red giant branch at the bump ($\sim$80,000 years, CB00).

By directly measuring the angular diameter of HD 148293, we are able to calculate its $T_{\rm eff}$ when combined with other observed quantities, such as interstellar absorption and bolometric corrections. We then modified the H-R diagram presented in CB00 to support their claim of proximity to the red-giant bump. Section 2 describes our observing procedure, Section 3 discusses how HD 148293's angular diameter and $T_{\rm eff}$ were determined, and Section 4 explores the physical implications of the new measurements.

%%%%%%%%%%%%%%%%%%%%%%%%% Interferometric observations %%%%%%%%%%%%%%%%%%%%%%%%%
\section{Interferometric observations}
Interferometric observations were obtained using the CHARA Array, a six element Y-shaped optical-infrared interferometer located on Mount Wilson, California \citep{2005ApJ...628..453T}. All observations used the pupil-plane ``CHARA Classic'' beam combiner in the $K'$-band at 2.14~$\mu$m while visible wavelengths (470-800 nm) were used for tracking and tip/tilt corrections. The observing procedure and data reduction process employed here are described in \citet{2005ApJ...628..439M}. We observed HD~148293 over two nights using two telescope pairs with different baseline lengths: 30 July 2010 using the E2-W2 pair with a baseline of approximately 156 m and 31 July 2010 using the W2-S2 pair at approximately 177 m.\footnote{The three arms of the CHARA Array are denoted by their cardinal directions: ``S'', ``E'', and ``W'' are south, east, and west, respectively. Each arm bears two telescopes, numbered ``1'' for the telescope farthest from the beam combining laboratory and ``2'' for the telescope closer to the lab. The ``baseline'' is the distance between the telescopes.} 

Two calibrators (HD 145454 and HD 147321) were selected to be single single stars with expected visibility amplitudes $>$95$\%$ so they were nearly unresolved on the baselines used, which meant uncertainties in the calibrator's diameter did not affect the target's diameter calculation as much as if the calibrator star had a significant angular size on the sky. We interleaved calibrator and target star observations so that every target was flanked by calibrator observations made as close in time as possible, which allowed us to convert instrumental target and calibrator visibilities to calibrated visibilities for the target. 

\begin{figure}[h]
\includegraphics[width=0.5\textwidth]{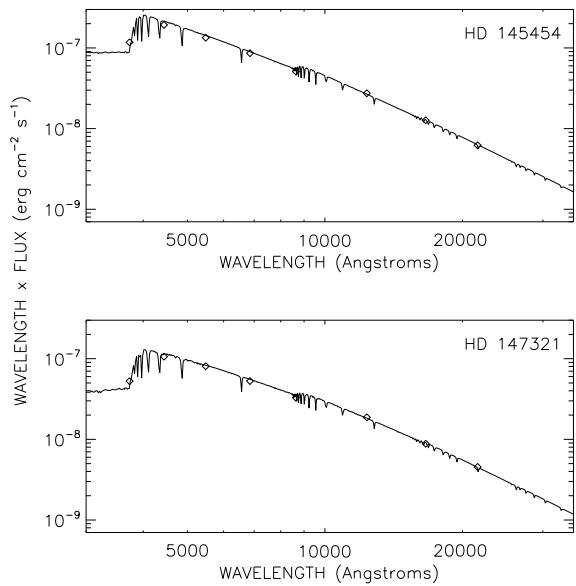}
\caption{SED fits for the calibrator stars HD 145454 and HD 147321. The diamonds are fluxes derived from $UBVRI JHK$ photometry (left to right) and the solid lines are the Kurucz stellar models of the stars. See Table \ref{calibrators} for the values used to create the fits.}
  \label{seds}
\end{figure}

To check for possible unseen close companions that would contaminate our observations, we created spectral energy distribution (SED) fits based on published $UBVRIJHK$ photometric values obtained from the literature for each calibrator to establish diameter estimates. We combined the photometry with Kurucz model atmospheres\footnote{Available to download at http://kurucz.cfa.harvard.edu.} based on $T_{\rm eff}$ and log~$g$ values to calculate limb-darkened angular diameters for the calibrators. The stellar models were fit to observed photometry after converting magnitudes to fluxes using \citet[][$UBVRI$]{1996AJ....112..307C} and \citet[][$JHK$]{2003AJ....126.1090C}. The photometry, $T_{\rm eff}$ and log~$g$ values, and resulting limb-darkened angular diameters for the calibrators are listed in Table \ref{calibrators}. There were no hints of excess emission associated with a low-mass stellar companion or circumstellar disk in the calibrators' SED fits (see Figure \ref{seds}).

\begin{deluxetable}{lccl}
\tablewidth{0pc}
%\tabletypesize{\scriptsize}
\tablecaption{Calibrator Information.\label{calibrators}}
\tablehead{ \colhead{Parameter} & \colhead{HD 145454} & \colhead{HD 147321} & \colhead{Source} }
\startdata
$U$ magnitude & 5.35 & 6.22 & \citet{Mermilliod} \\
$B$ magnitude & 5.42 & 6.07 & \citet{Mermilliod} \\
$V$ magnitude & 5.44 & 5.99 & \citet{Mermilliod} \\
$R$ magnitude & 5.46 & 5.99 & \citet{2003AJ....125..984M} \\
$I$ magnitude & 5.50 & 5.98 & \citet{2003AJ....125..984M} \\
$J$ magnitude & 5.37 & 5.79 & \citet{2003tmc..book.....C} \\
$H$ magnitude & 5.43 & 5.82 & \citet{2003tmc..book.....C} \\
$K$ magnitude & 5.43 & 5.77 & \citet{2003tmc..book.....C} \\
$T_{\rm eff}$ (K) & 9772 &  & \citet{1999AandA...352..555A} \\
log $g$ (cm s$^{-2}$) & 4.13 &  & \citet{1999AandA...352..555A} \\
$T_{\rm eff}$ (K) & & 8600 & \citet{2010yCat.2300....0L} \\
log $g$ (cm s$^{-2}$) & & 4.2 & \citet{2010yCat.2300....0L} \\
$\theta_{\rm LD}$ (mas) & 0.268$\; \pm \;$0.015 & 0.240$\; \pm \;$0.010 & \\
\enddata
\end{deluxetable}

%%%%%%%%%%%%%%%%%%%%%%%% Angular diameter determinations %%%%%%%%%%%%%%%%%%%%%%%%
\section{Determination of angular diameter and $T_{\rm eff}$}
The observed quantity of an interferometer is defined as the visibility ($V$), which is fit to a model of a uniformly-illuminated disk (UD) that represents the observed face of the star. Diameter fits to $V$ were based upon the UD approximation given by $V = 2 J_1(x) / x$, where $J_1$ is the first-order Bessel function and $x = \pi B \theta_{\rm UD} \lambda^{-1}$, where $B$ is the projected baseline at the star's position, $\theta_{\rm UD}$ is the apparent UD angular diameter of the star, and $\lambda$ is the effective wavelength of the observation \citep{1992ARAandA..30..457S}. A more realistic model of a star's disk involves limb-darkening (LD), and relationship incorporating the linear LD coefficient $\mu_{\lambda}$ \citep{1974MNRAS.167..475H} is:
\begin{equation}\begin{split}
V = \left( {1-\mu_\lambda \over 2} + {\mu_\lambda \over 3} \right)^{-2}
\times \\
\left[(1-\mu_\lambda) {J_1(\rm x) \over \rm x} + \mu_\lambda {\left( \frac{\pi}{2} \right)^{1/2} \frac{J_{3/2}(\rm x)}{\rm x^{3/2}}} \right] .
\end{split}\end{equation}
Table \ref{calib_visy} lists the Modified Julian Date (MJD), projected baseline ($B$) at the time of observation, projected baseline position angle ($\Theta$), calibrated visibility ($V$), and error in $V$ ($\sigma V$) for HD 148293. 

\begin{deluxetable}{cccccc}
\tablewidth{0pc}
\tablecaption{HD 148293's Calibrated Visibilities.\label{calib_visy}}

\tablehead{\colhead{Calib} &  \colhead{ } & \colhead{$B$} & \colhead{$\Theta$} & \colhead{ } & \colhead{ }
 \\
\colhead{HD} & \colhead{MJD} & \colhead{(m)} & \colhead{(deg)} & \colhead{$V$} & \colhead{$\sigma V$} \\ }
\startdata
147321 & 55407.165 & 210.93 & 251.2 & 0.518 & 0.077 \\
 & 55407.174 & 211.75 & 253.7 & 0.441 & 0.068 \\
 & 55407.182 & 212.40 & 255.9 & 0.465 & 0.063 \\
 & 55407.191 & 212.95 & 258.2 & 0.439 & 0.061 \\
 & 55407.200 & 213.41 & 260.4 & 0.542 & 0.062 \\ 
 & 55407.208 & 213.77 & 262.7 & 0.504 & 0.057 \\ 
 & 55407.217 & 214.03 & 264.9 & 0.426 & 0.055 \\
145454 & 55408.275 & 156.09 & 243.2 & 0.759  & 0.072 \\
 & 55408.283 & 156.15 & 246.0 & 0.768 & 0.061 \\
 & 55408.291 & 156.18 & 248.7 & 0.752 & 0.035 \\
 & 55408.304 & 156.22 & 252.8 & 0.750 & 0.057 \\
 & 55408.312 & 156.24 & 255.7 & 0.765 & 0.080 \\
147321 & 55408.275 & 156.09 & 243.2 & 0.685 & 0.045 \\
 & 55408.283 & 156.15 & 246.0 & 0.728 & 0.052 \\
 & 55408.291 & 156.18 & 248.7 & 0.730 & 0.033 \\
 & 55408.304 & 156.22 & 252.8 & 0.763 & 0.052 \\
 & 55408.312 & 156.24 & 255.7 & 0.747 & 0.072 \\
\enddata
\tablecomments{The projected baseline position angle ($\Theta$) is calculated to be east of north.}
\end{deluxetable}

The LD coefficient was obtained from \citet{1995AandAS..114..247C} after adopting the $T_{\rm eff}$ and log~$g$ values for HD 148293 and the resulting LD angular diameter is listed in Table \ref{parameters}. The errors on the UD and LD diameters are 4$\%$ and the difference between the UD and LD diameters is on the order of a few percent, and the final angular diameter is little affected by the choice of $\mu_{\lambda}$\footnote{A 20$\%$ change in the $\mu_{\lambda}$ leads to a change in the measured LD diameter is less than $1\%$}. Additionally, the combination of the interferometric measurement of the star's angular diameter plus the \emph{Hipparcos} parallax \citep{2007hnrr.book.....V} allowed us to determine the star's physical radius. This result is also listed in Table \ref{parameters}.

\begin{deluxetable}{lcl}
\tablewidth{0pc}
%\tabletypesize{\scriptsize}
\tablecaption{HD 148293 Stellar Parameters.\label{parameters}}
\tablehead{ \colhead{Parameter} & \colhead{Value} & \colhead{Reference} }
\startdata
[Fe/H]            & +0.08 & \citet{1997AandAS..124..299C} \\
$V$ magnitude     & 5.25 & \citet{Mermilliod} \\
$K$ magnitude     & 2.83$\; \pm \;$0.11 & \citet{1969tmss.book.....N} \\
$A_{\rm V}$       & 0.04 & \citet{2005AandA...430..165F} \\
BC                & 0.36$\; \pm \;$0.10 & \citet{1999AandAS..140..261A} \\
%Luminosity ($L_\odot$) & 73.0$\; \pm \;$7.0  & Calculated here \\
$F_{\rm BOL}$ (10$^{-8}$ erg s$^{-1}$ cm$^{-2}$) & 28.9$\; \pm \;$2.8 & Calculated here \\
$\theta_{\rm UD}$ (mas) & 1.439$\; \pm \;$0.059 (4$\%$) & Measured here \\
$\theta_{\rm LD}$ (mas) & 1.480$\; \pm \;$0.060 (4$\%$) & Measured here \\
$R_{\rm linear}$ ($R_\odot$) &  14.3$\; \pm \;$0.6 (4$\%$) & Measured here \\
$T_{\rm eff}$ (K) & 4640$\; \pm \;$100 & \citet{1989ApJS...71..293B} \\
$T_{\rm eff}$ (K) & 4460$\; \pm \;$141 (3$\%$) & Measured here \\
\enddata
%\tablecomments{}
\end{deluxetable}

\begin{figure}[h]
\includegraphics[width=0.35\textwidth, angle=90]{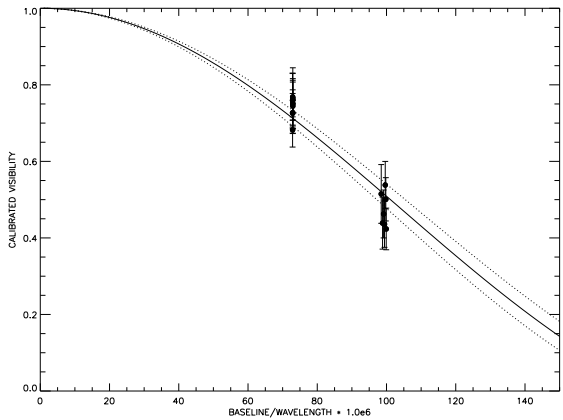}
\caption{HD 148293 LD disk diameter fit. The solid line represents the theoretical visibility curve for a star with the best fit $\theta_{\rm LD}$, the dotted lines are the 1$\sigma$ error limits of the diameter fit, the filled circles are the calibrated visibilities, and the vertical lines are the measured errors.}
  \label{hd148293}
\end{figure}

For the $\theta_{\rm LD}$ fit, the errors were derived via the reduced $\chi^2$ minimization method \citep{2003psa..book.....W,1992nrca.book.....P}: the diameter fit with the lowest $\chi^2$ was found and the corresponding diameter was the final $\theta_{\rm LD}$ for the star. The errors were calculated by finding the diameter at $\chi^2 + 1$ on either side of the minimum $\chi^2$ and determining the difference between the $\chi^2$ diameter and $\chi^2 +1$ diameter. In calculating the diameter errors in Table \ref{parameters}, we adjusted the estimated visibility errors to force the reduced $\chi^2$ to unity because when this is omitted, the reduced $\chi^2$ is well under 1.0, indicating we are overestimating the errors in our calibrated visibilities. Figure \ref{hd148293} shows the LD diameter fit for HD 148293. Though several points lie outside the 1-$\sigma$ line of the diameter fit, the errors in the individual visibility points overlap the diameter fit itself.

Once $\theta_{\rm LD}$ was determined interferometrically, the $T_{\rm eff}$ was calculated using the relation 
\begin{equation}
F_{\rm BOL} = {1 \over 4} \theta_{\rm LD}^2 \sigma T_{\rm eff}^4,
\end{equation}
where $F_{\rm BOL}$ is the bolometric flux and $\sigma$ is the Stefan-Bolzmann constant. $F_{\rm BOL}$ was determined in the following way: HD 148293's $V$ and $K$ magnitudes were dereddened using the extinction curve described in \citet{1989ApJ...345..245C} and its interstellar absorption ($A_{\rm V}$) value was from \citet{2005AandA...430..165F}. The intrinsic broad-band color ($V-K$) was calculated and the bolometric correction (BC) was determined by interpolating between the [Fe/H] = 0.0 and +0.2 tables from \citet{1999AandAS..140..261A}. They point out that in the range of 6000 K $\geq T_{\rm eff} \geq$ 4000 K, their BC calibration is symmetrically distributed around a $\pm$0.10 magnitude band when compared to other calibrations, so we assigned the BC an error of 0.10. The bolometric flux was then determined by applying the BC and the $T_{\rm eff}$ was calculated. See Table \ref{parameters} for a summary of these parameters.

%%%%%%%%%%%%%%%%%%%%%%%%%%%%% Results %%%%%%%%%%%%%%%%%%%%%%%%%%%%%
\section{Results and discussion}

\begin{figure}[h]
\includegraphics[width=0.35\textwidth, angle=90]{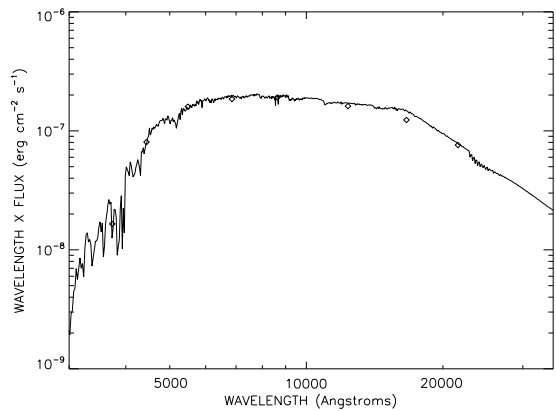}
\caption{HD 148293 SED fit. The diamonds are fluxes derived from $UBVRI JHK$ photometry (left to right) and the solid line is the Kurucz stellar model of a star with $T_{\rm eff}$ = 4571 K and log $g$ = 2.41 from \citet{1999AandA...352..555A}. There were no quoted errors for the $UBVRI$ measurements, $\sigma_J \sim 7\%$, $\sigma_H \sim 7$, and $\sigma_K \sim 12\%$, which are not indicated on the plot.}
  \label{sed}
\end{figure}

As a check to our measured diameter, limb-darkened angular diameters were estimated using two additional methods: (1) by producing an SED fit (see Figure \ref{sed}) as described in Section 2, where $UBV$ photometry is from \citet{Mermilliod}, $RI$ photometry is from \citet{2003AJ....125..984M}; and $JHK$ photometry is from \citet{2003tmc..book.....C}; and (2) using the relationship described in \citet{2004AandA...426..297K} between the ($V-K$) color and log $\theta_{\rm LD}$. Our measured $\theta_{\rm LD}$ is 1.480$\; \pm \;$0.060 mas, the SED fit estimates 1.418$\; \pm \;$0.085 mas, and the color-diameter relationship produces 1.465$\; \pm \;$0.692 mas. Because 2MASS measurements saturate at magnitudes brighter than $\sim$3.5 in the $K$-band even when using the shortest exposure time\footnote{Explanatory Supplement to the 2MASS All Sky Data Release and Extended Mission Products, http://www.ipac.caltech.edu/2mass/releases/allsky/doc/.}, we used the $K$ magnitude from the Two-Micron Sky Survey \citep{1969tmss.book.....N} for the color-diameter determination. 

The main sources of errors for the three methods are uncertainties in visibility measurements for the interferometric measurement, uncertainties in the comparison between observed fluxes and the model fluxes for a given $T_{\rm eff}$ and log~$g$ for the SED estimate, and uncertainties in the $K$ magnitude for the color-diameter determination. All three diameters agree within their errors but our interferometric measurements provide an error approximately 40$\%$ and 12$\times$ smaller than the latter two methods, respectively.

We also wanted to compare our $T_{\rm eff}$ with those obtained from the literature (see Table \ref{temps}). Values range from 4390 K \citep{2010yCat.2300....0L} to 4693$\pm$24 K \citep{2011AandA...525A..71W}. The methods are varied and are based on spectroscopic observations, photometric measurements, spectral typing, color-$T_{\rm eff}$ relationships, the correlation between absorption line features and stellar parameters, and the star's position on the color-magnitude diagram. Most temperatures do not have errors indicated in the source papers. Our measurement of 4460$\pm$141 K is on the cooler side of the values from the literature, though the entire range spans only $\sim$300 K.

\begin{deluxetable}{lll}
\tablewidth{0pc}
%\tabletypesize{\scriptsize}
\tablecaption{HD 148293 $T_{\rm eff}$ from the Literature.\label{temps}}
\tablehead{ \colhead{$T_{\rm eff}$ (K)} & \colhead{Reference} & \colhead{Method Used} }
\startdata
4390 & \citet{2010yCat.2300....0L} & Based on spectral type \\
4420 & \citet{2008AstL...34..785G} & Photometric measurements \\
4420 & \citet{2003AJ....125..359W} & Based on spectral type \\
4560 & \citet{1999AandAS..140..261A} & ($B-V$) - $T_{\rm eff}$ relationship \\
4571$\pm$12 & \citet{1999AandA...352..555A} & Position on color-magnitude diagram \\
4585$^{+287}_{-189}$ & \citet{2006ApJ...638.1004A} & Photometric measurements \\
4640$\pm$100 & \citet{1989ApJS...71..293B} & Photometry and color-$T_{\rm eff}$ relations \\
4640 & \citet{2007ApJS..171..146S} & Relation between absorption line features \\
     &                             & and stellar parameters \\
4650 & \citet{1990ApJS...74.1075M} & Color - $T_{\rm eff}$ relationship \\
4650 & \citet{1997AandAS..124..299C, 2001AandA...373..159C} & Spectroscopic measurements \\
4693$\pm$24 & \citet{2011AandA...525A..71W} & Spectroscopic measurements \\
4460$\pm$141 & Measured here & Interferometric measurement \\
\enddata
\end{deluxetable}

\begin{figure}[h]
\includegraphics[width=0.35\textwidth, angle=90]{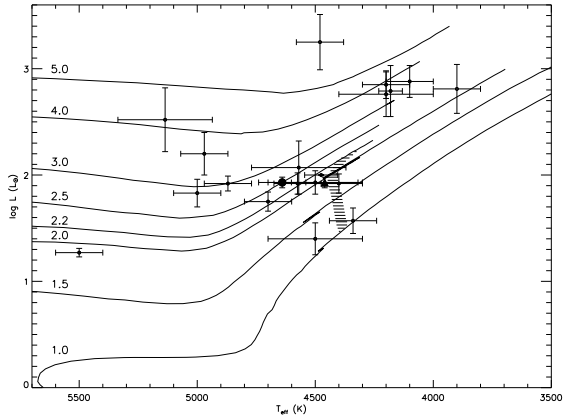}
\caption{H-R diagram for the Li-rich giants. The small circles are the stars from CB00, the large circle is their original placement of HD 148293, and the large triangle is our new placement. The evolutionary tracks are marked by the stars' masses, and the shaded region indicates the red giant bump region. The luminosity and temperature error bars for the Charbonnel \& Balachandran points are from uncertainties in the Hipparcos parallaxes and from the literature, respectively. Our error bars are based on interferometric observations, uncertainties in Hipparcos parallaxes, and bolometric corrections.}
  \label{hr}
\end{figure}

With our newly calculated $T_{\rm eff}$, we are able to place HD 148293 on the H-R diagram from CB00. Figure \ref{hr} shows a reproduction of their Figure 1 with Li-rich giant stars plotted along with evolutionary tracks for a range of stellar masses. We used the YZVAR stellar model \citep{2008AandA...484..815B, 2009AandA...508..355B} for HD 148293's metallicity of +0.08 \citep{1997AandAS..124..299C}.

CB00 hypothesized that HD 148293 is at the red-giant bump, even though the $T_{\rm eff}$ they used from \citet{1989ApJS...71..293B} places the star slightly to the left of the bump. They claim the temperature shift required to place HD 148293 in the red-giant bump is not unreasonable ($\sim$200 K), given the uncertainties in the $T_{\rm eff}$ measurement. Our $T_{\rm eff}$ is 180 K cooler than the value they used and places HD 148293 closer to and within the error bar of the red-giant bump, supporting their hypothesis. CB00 do not seem to believe that the slight shift in $T_{\rm eff}$ significantly affects the Li abundance calculated by Brown et al., so it would appear likely that we are seeing HD 148293 during a very brief stage in its evolutionary process where Li is being produced or was produced very recently.

\acknowledgements

The CHARA Array is funded by the National Science Foundation through NSF grant AST-0606958 and by Georgia State University through the College of Arts and Sciences, and by the W.M. Keck Foundation. STR acknowledges partial support by NASA grant NNH09AK731. This research has made use of the SIMBAD database, operated at CDS, Strasbourg, France.  This publication makes use of data products from the Two Micron All Sky Survey, which is a joint project of the University of Massachusetts and the Infrared Processing and Analysis Center/California Institute of Technology, funded by the National Aeronautics and Space Administration and the National Science Foundation.

\end{document}